\documentclass[letter,twocolumn]{jpsj3}
\usepackage{graphicx,amsmath,amssymb,bm}

\allowdisplaybreaks 

\usepackage{color}
\definecolor{Green}{rgb}{0,0.7,0}
\newcommand{\bk}{\bm{k}}

\newcommand{\bq}{\bm{q}}
\newcommand{\bkD}{\bk_{\rm D}}

\newcommand{\ep}{\varepsilon}

\newcommand{\stf}{$\alpha$-STF$_2$I$_3$}
\newcommand{\ET}{$\alpha$-ET$_2$I$_3$}
\newcommand{\stfb}{$\alpha$-STF$_2$I$_3$ }

\newcommand{\bets}{$\alpha$-BETS$_2$I$_3$}

\begin{document}

%\today

%-------------------------------
\title{
Exotic  Dirac Cones  on the Band Structure of $\alpha$-STF$_2$I$_3$  
 at Ambient Temperature and Pressure
}
\author{
Toshio \textsc{Naito}$^{1}$
\thanks{E-mail: tnaito@ehime-u.ac.jp},
Ryusei Doi$^{1}$, and 
Yoshikazu \textsc{Suzumura}$^{2}$
}
\inst{
$^1$Department of Chemistry, Ehime University, Matsuyama 790-8577, Japan
 \\
$^2$Department of Physics, Nagoya University, 
 Nagoya 464-8602, Japan \\
}

%\recdate{\today}
%-------------------------------
\abst{
The quasi-two-dimensional molecular semimetallic conductor \stfb is isostructural
with \ET. The latter possesses a unique band structure showing a zero-gap state with Dirac cones under high pressure, whereas the band 
structure of the former has been elusive because of  heavy
 disorder at all the donor sites.
 To elucidate the band structure of \stf, a theoretical method  based on the observed atomic parameters at 296 K and 1 bar  has been  proposed. 
 The results suggest that  the STF salt should  have  a band structure with Dirac cones under ambient pressure 
 and temperature, which should promote 
 future experimental studies on this system. 
Using the extended H\"uckel method, we demonstrate that the Dirac points of \stf \;  are  aligned  to be symmetric  with respect to a time reversal invariant momentum (TRIM). 
  Such novel Dirac cones, where the energy difference between the conduction and valence bands has considerable anisotropy, are clarified in terms of 
 the parity of the wavefunction at the TRIM. 
We propose the conductivity measurement on the present \stf,  which is expected to show a large anisotropy as a characteristic of the present  Dirac electrons.
}

%\kword{parity, Dirac point, $\alpha$-(BEDT-TTF)$_2$I$_3$, \stf
%Dirac cone, merging}

%\begin{document}

\maketitle

In a recent development in  organic conductors,\cite{Seo2004,Kajita_JPSJ2014}
 Dirac electrons have been studied extensively 
since the zero-gap state with Dirac cones~\cite{a6} has been found 
 in \ET\; (ET = BEDT-TTF =
bis(ethylenedithio)tetrathiafulvalene) under high pressure ($>$1.5 GPa)
using a tight-binding model with transfer integrals.\cite{Kajita_JPSJ2014,Kondo2005_RSI76,a7,a8,a11} 
\ET\; 
 is isostructural with  
\bets\cite{TN1} 
and \stf\; as 
 the $\alpha$ salt,\cite{a1,a2,a3,a4} 
%-------------------------------------
(BETS = BEDT-TSF = bis(ethylenedithio)tetraselenafulvalene).
They share a three-quarter-filled band and four molecules per unit cell.
The organic donor STF,~\cite{a1,a2,a3,a4}
 where STF is 
bis(ethylenedithio)diselenadithiafulvalene, 
is an analogue of ET and BETS, 
has produced a number of salts isostructural with those of ET and BETS, including
$\alpha$-D$_2$I$_3$ (D = ET, STF, and  BETS) (Fig.~1).

\begin{figure}
  \centering
\includegraphics[width=8cm]{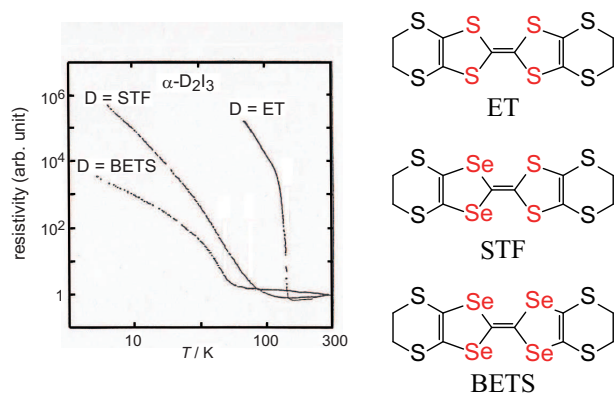}   
  \caption{(Color online)
Electrical behavior of $\alpha$-D$_2$I$_3$~\cite{a2}
 (D = ET, STF, and BETS) with the molecular structures of donor D.
}
\label{fig1}
\end{figure}

The most important difference between STF and the  other organic donors (ET and BETS)
 lies in the molecular symmetry. Because of the lower symmetry, all the known STF salts include orientational disorder at the donor sites in the crystals. 
However, regarding electrical, magnetic, and optical properties in the solid states,  many of the STF salts exhibit intermediate behavior between the isostructural ET and BETS salts, as if the solids do not contain any disorder but consist of a symmetrical donor containing imaginary atoms between selenium and sulfur at the inner chalcogen atoms.~\cite{a1,a2,a3,a4}
%\cite{a1,a2,a3,a4.a5} 
The observation 
  suggests that STF salts should lie between the isostructural ET and BETS salts with stronger/weaker intermolecular interactions among donor molecules than the ET/BETS salts.
A typical example is $\alpha$-D$_2$I$_3$ (D = ET, STF, BETS), where one can recognize systematic variation in the electrical behavior 
in the three salts as shown in Fig.~\ref{fig1}.~\cite{a2}

%========  Fig 1 ================
%--------------------------------------

For \bets, the  temperature dependence of the spin susceptibility\cite{Takahashi2010_JPSJ,a5}
 shows a marked decrease  even in the normal state, 
suggesting an unusual  electronic state ascribed to the peculiar energy band. 
 The band calculation indicates that there should be a Dirac point 
 with overtilted cones, and that \bets\; should remain metallic.~\cite{Kondo2009_JPSJ} The zero-gap state  is studied by adding a site potential~\cite{Kondo2009_JPSJ} or 
     taking account of the correlation,~\cite{Morinari_Suzumura} where 
 the latter shows the Dirac point near the insulating state 
 followed by the merging of Dirac points  at one of 
  the time reversal invariant momenta (TRIMs).
In this context, the mechanism to produce Dirac points 
  should be more complicated  in  \bets\; than in \ET, 
 although a Dirac point by itself originates from the intrinsic property
  of band crossing.    
Although \stf\;  is also expected to show such an electronic state 
% due to molecular structure facing  the other two salts, 
owing to the close similarities in  the molecular structures, 
  the band structure is  yet to be clarified.

In this Letter, we demonstrate the zero-gap state with 
  the Dirac cone in \stfb 
by a modified method of estimation of transfer integrals,
which is based on single-crystal X-ray structural analysis 
 [Fig.~2(a)].\cite{a2}
The paper is organized as follows. First, we 
 propose a tentative new method  for \stf\;  
 to obtain the transfer integrals for the tight-binding model.
Next, on the basis of this model,
  we show   theoretically the zero-gap state with 
   Dirac cones, where the Dirac points are aligned  toward  a TRIM. 

%----------------------
 In the band calculation of \stf, 
 the most difficult problem is to satisfy the crystallographic requirements in symmetries, which could not be done by locating unsymmetrical donor wavefunctions at the STF sites. 
To calculate the band structure,  
 it is crucial to retain the crystallographic symmetries required for both a particular donor site and the entire 
 donor arrangement.
%========  Fig 2 ================
\begin{figure}
  \centering
\includegraphics[width=4.5cm]{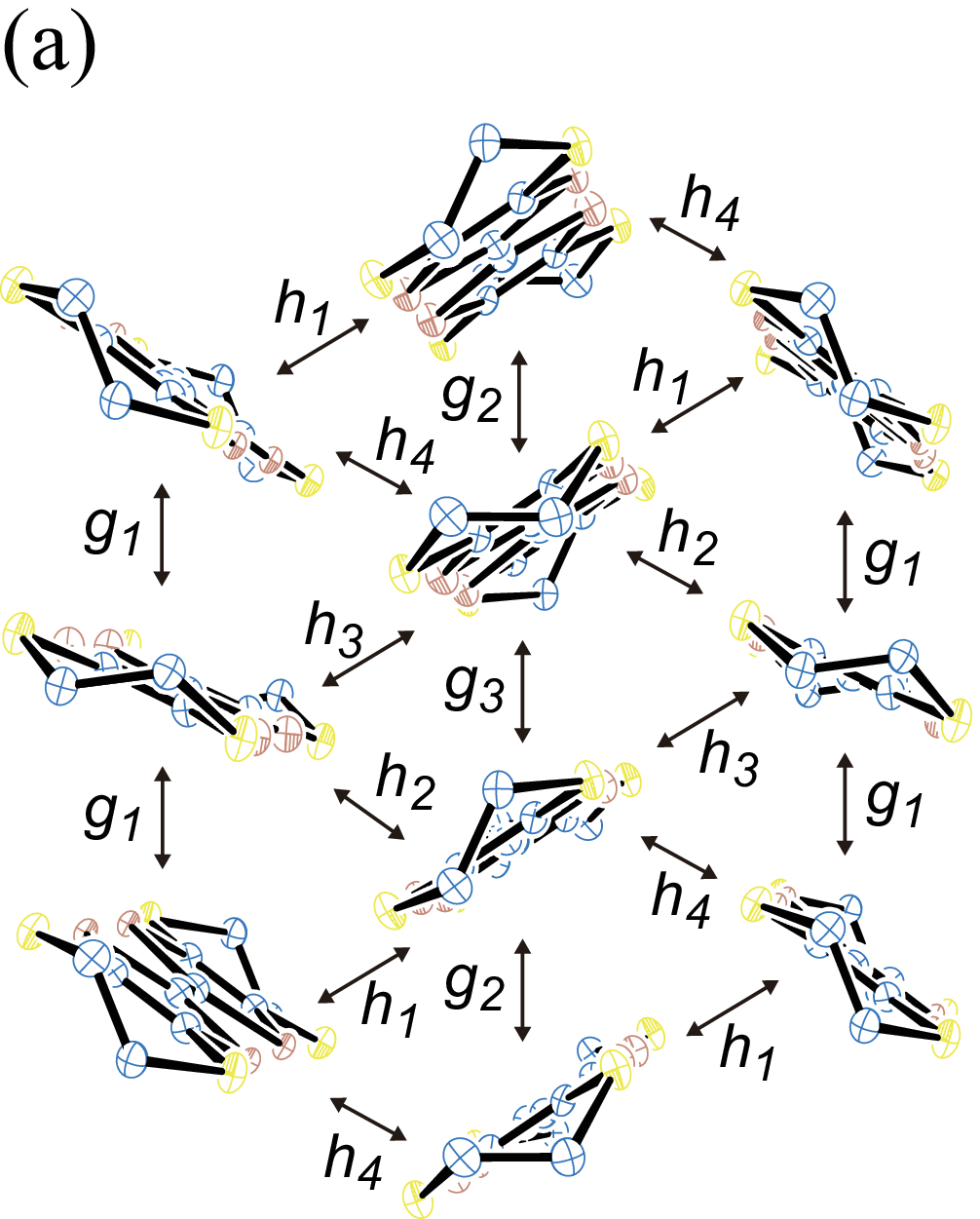}   
\includegraphics[width=5.5cm]{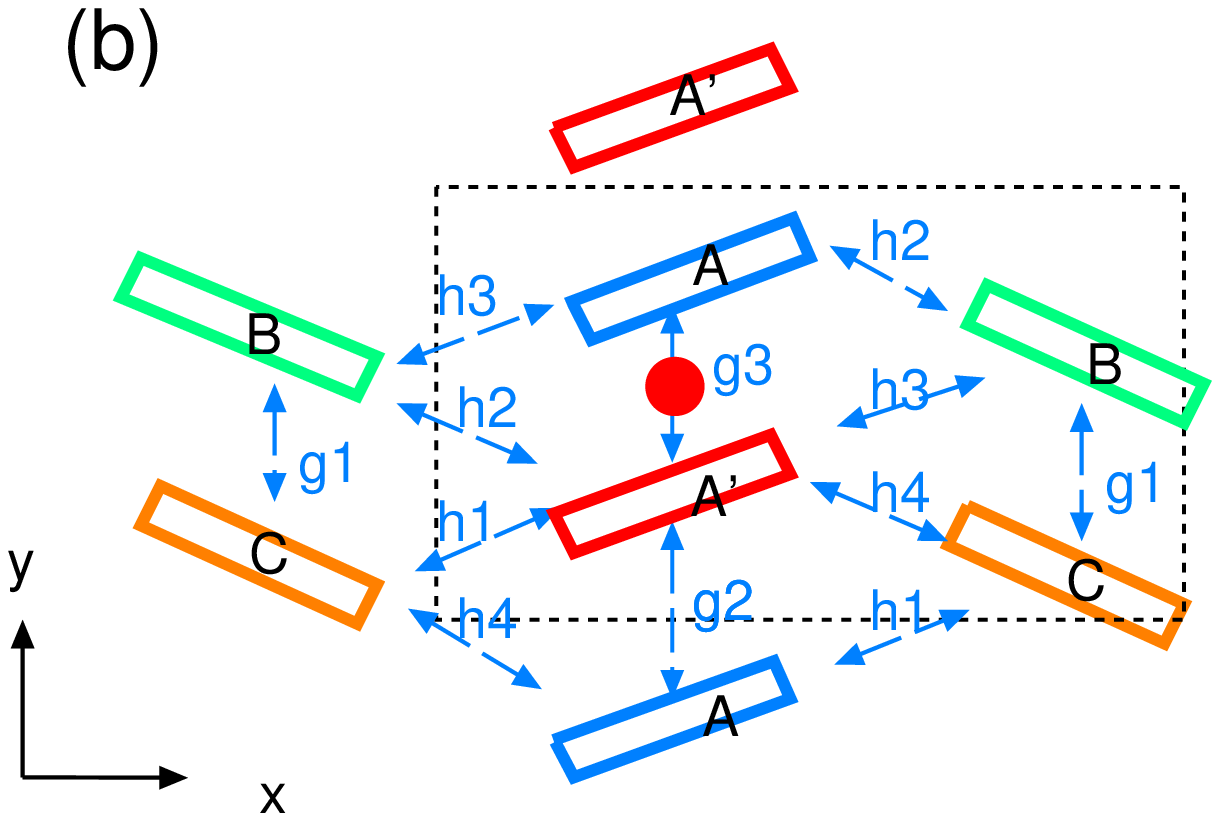}   
  \caption{(Color online)
 Molecular arrangement of STF in \stf. (a) Intermolecular interactions in the conduction sheet. Brown spheres indicate  disordered chalcogen atoms (Se or S), which are included in every STF molecule.
 Arrows indicated by $g_i$ ($i$ = 1--3) and $h_j$ ($j$ = 1--4) show  
  intermolecular interactions. Hydrogen atoms are omitted for clarity. 
(b) Schematic picture of (a). A, B, and C indicate crystallographically independent STF molecules, while A and A' are interrelated by an 
 inversion center (closed circle). Four molecules in the same unit cell (dotted line) have 
 a common phase of wavefunctions. 
}
%\label{fig:structure}
\label{fig2}
\end{figure}
%--------------------------------------

%========  Fig 3 ================
\begin{figure}
  \centering
\includegraphics[width=8cm]{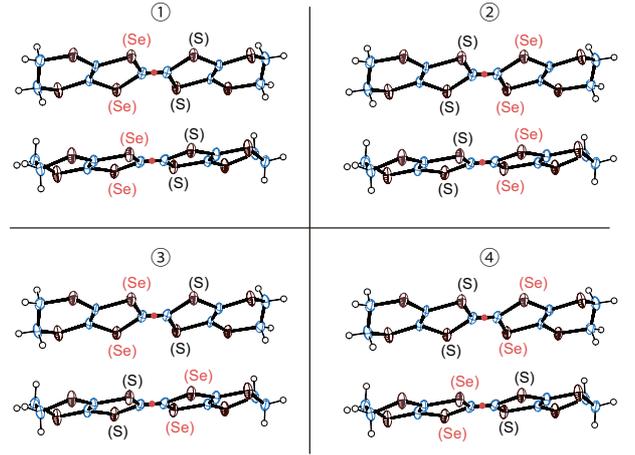}   
  \caption{(Color online)
Four patterns of possible arrangements of an interacting pair 
of STF molecules considered in the calculation, 
 where 
$\textcircled{1}$, 
$\textcircled{2}$, 
$\textcircled{3}$, and 
$\textcircled{4}$ 
  correspond  to those in Table S4 in Ref.~\citen{a12}.
All these patterns occur in the crystal with an equal probability to 
each other to reproduce the observed structure.
}
\label{fig3}
\end{figure}
%--------------------------------------

\begin{figure}
  \centering
\includegraphics[width=5cm]{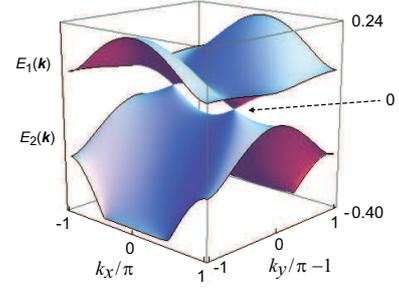}  
  \caption{(Color online)
 Energy bands of $E_1({\bk})$ and  $E_2({\bk})$ 
 as  functions of $\bk$.   The  Dirac points exist   at   
 $ \bkD = (k_x/\pi, k_y/\pi-1)= \pm (0.61,-0.18)$. 
 The relation  $E_1(\bk) \ge E_1(\bkD)= E_2(\bkD) \ge E_2(\bk)$ 
 suggests  the zero-gap state.
 The  energy at  the Dirac point  
is equal to the chemical potential.
}
\label{fig4}
\end{figure}

\begin{figure}
  \centering
\includegraphics[width=7cm]{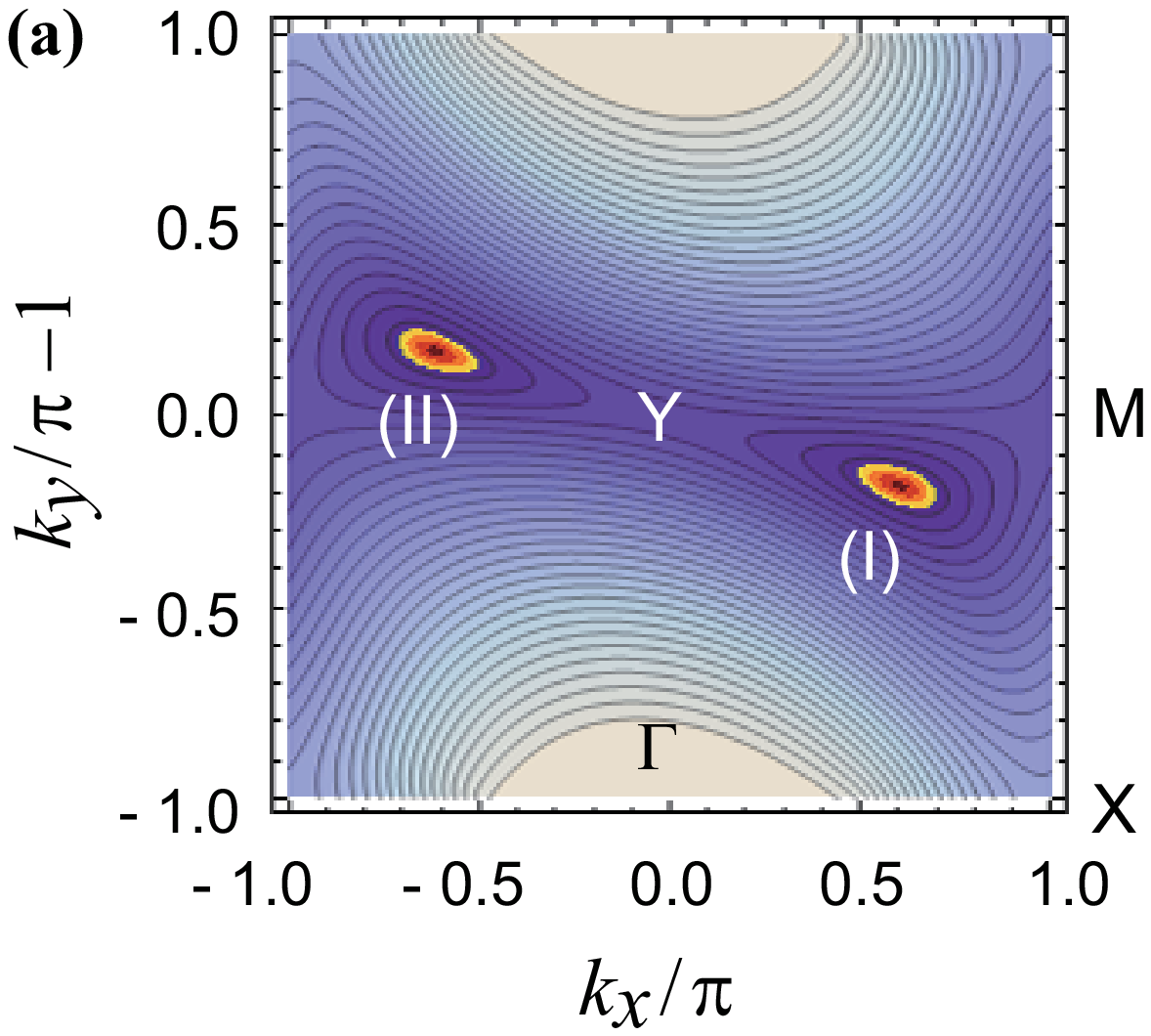}   
\includegraphics[width=6cm]{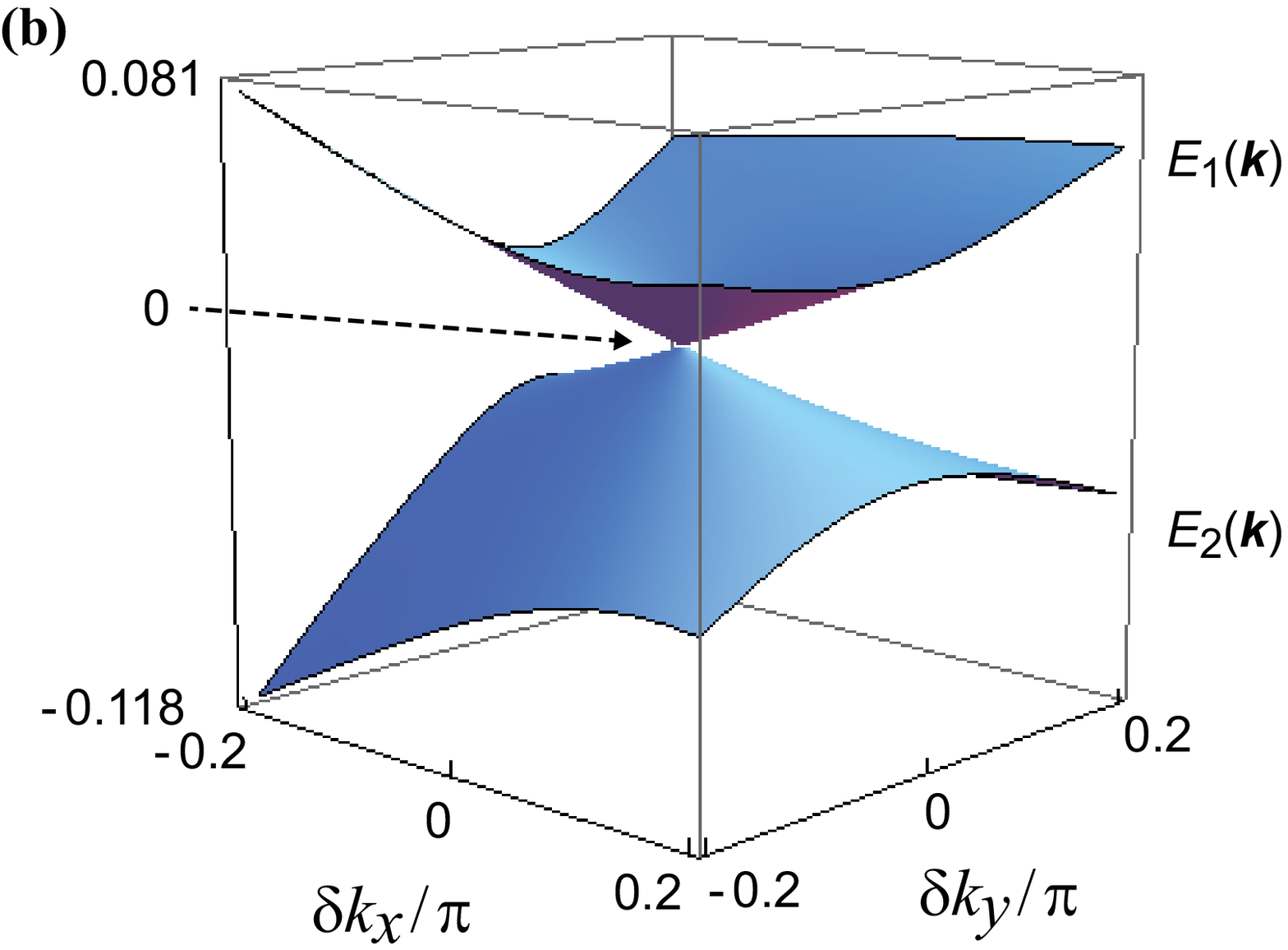}  \\ 
  \caption{(Color online)
(a) Contour plot of $E_1(\bk)$ - $E_2(\bk)$.
 The orange region denotes $0 < E_1(\bk) - E_2(\bk) < 0.03$,
 where the Dirac points 
 with $\bkD/\pi -(0,1) = (k_x/\pi, k_y/\pi-1)$ = (0.61, -0.18) (I),  and 
 (-0.61, 0.18) (II) 
are seen at the center of  each orange  region. 
(b) Dirac cones for the Dirac point (I) in (a) on the plane of 
  $-0.2< \delta k_x/\pi < 0.2$ and $-0.2 < \delta k_y/\pi <0.2$. 
 $\delta \bk = \bk - \bkD$.  
$\bkD/\pi - (0,1) = (0.61, -0.18)$.
}
\label{fig5}
\end{figure}

In this study, we have
noticed that all the occupancies at the inner chalcogen atoms are 50:50 = S:Se in the STF salt (Tables S1 and S2 in Ref.~\citen{a12}). 
Thus,   
 we  carried out a band structure calculation, assuming that the effective 
%------------------
 transfer integrals ($t_{\rm eff}$)  between disordered STF molecules should
 be treated approximately  using the averaged values of 
all the possible orientations, i.e., four patterns of molecular arrangements in each pair of interacting STF molecules (Fig.~\ref{fig3}).
%    with 
% an  inversion center located  at  the middle of the molecule. 
The  grounds for this assumption are obtained from the single-crystal X-ray structural analysis on \stf\cite{a12}
 [Fig.~2(a)]\cite{a2}.
 The asymmetric unit contains two STF molecules in total: an entire STF at 
 a general position plus two halves of STF on the inversion centers, all of which are disordered in the molecular orientation (Fig.~3).
%-----------------
The occupancies of S/Se in the inner chalcogen atoms of STF are all 50\%/50\%. 
This means that the electron densities of the inner four chalcogen atoms in every STF molecule   are averaged to satisfy the inversion symmetry.
The obtained transfer  integrals corresponding to 
% ($S_{\rm eff} \times 10^3$) are as follows;
 $g_1, g_2, g_3, h_1, h_2, h_3$, and $h_4$ in Figs.~\ref{fig2}(a) and 
 \ref{fig2}(b) are summarized in Table S4 (Model A) in Ref.~\citen{a12}
 based on the H\"uckel parameters in Table S3 in Ref.~\citen{a12}.
%-----------------------
We also examined the parameter dependence of the results, and have confirmed 
 that a different set of atomic parameters obtained from independent 
 structural analysis also gave qualitatively the same results: 
 a band structure with Dirac cones [Tables S2 and S4 (Model B), and  
Figs. S1(a) -- S1(c)].~\cite{a12} 
As shown in Figs.~\ref{fig4}, \ref{fig5}(a), and \ref{fig5}(b), 
the calculation 
 suggested that \stfb should have  a band structure 
 characterized  by its Dirac cones at 296 K and 1 bar.
The tight-binding band structure obtained on the basis of these transfer  integrals  qualitatively accounts for the electrical and magnetic properties of \stf, which should also be further examined  experimentally.

%-----------------
 Here, we make a brief comment on the effect of disorder on the band 
 structure and electronic properties. 
%----------------
 The disorder should affect the intrinsic electronic properties of Dirac cones 
 and those of a zero-gap semiconductor in \stf, which limits ourselves to 
comparing the calculated band structure with experimental results. 
Thus, it is 
 important to estimate the energy scale of disorder effects to clarify 
 the energy and temperature ranges where the band structure will be 
 directly manifested in the electronic properties. 
In Fig.~\ref{fig1}, \stfb exhibits nearly temperature-independent electrical 
behavior at $T \ge$100 K,
which is the  behavior characteristic   to a zero-gap 
 semiconductor having Dirac cones. Below $\sim$100 K, the resistivity 
 increases with decreasing temperature.
 Thus, our calculation results can be compared with the observed resistivity    
 behavior at $T \geq$ 100 K, and the disorder effect should overwhelm 
the intrinsic behavior below $\sim$100 K.
¡¡

%---------------------------------------------------------
Now, we examine theoretically  such an  exotic band structure 
in  \stfb  with  inversion symmetry between A and A', which is 
crucial for the existence of  the Dirac point.~\cite{Suzumura2016_JPSJ} 
% which  has been claimed  by a simple model.~\cite{Montambaux2009_EPJB72} 
This mechanism  is analyzed in terms of the parity inversion properties of energy bands  at the TRIM 
($\bm{G}/2$ with $\bm{G}$ being a reciprocal lattice vector).\cite{Piechon2013}

From   Ref.~\citen{a12},  
 the  transfer energies  (Model A) for a  tight-binding model 
 with   a square lattice 
  are obtained as   
  $g_1$ =0.0015,
  $g_2$ = 0.1420,
  $g_3$ = 0.0450,
  $h_1$ = -0.2767,
  $h_2$ = -0.2847,
  $h_3$ = 0.0092, and 
  $h_4$ = 0.0057,  
    in the unit of eV [Fig.~\ref{fig2}(b)].
The Hamiltonian is expressed as  
%----------- (1)------------------
\begin{eqnarray}
H = \sum_{i,j = 1}^N \sum_{\alpha, \beta = 1}^4
 t_{i,j; \alpha,\beta} a^{\dagger}_{i,\alpha} a_{j, \beta} \; , 
\label{eq:Hij}
\end{eqnarray}
where $a^{\dagger}_{i, \alpha}$ denotes a creation operator 
 of an electron 
 of molecule $\alpha$ 
 [A(1), A'(2), B(3), and C(4)]  in the unit cell 
  at  the $i$th lattice site.  
The transfer energies, $g_1, \cdots, h_4$ are given by  
  $t_{i,j; \alpha,\beta}$.
By using the Fourier transform 
 $a_j = 1/N^{1/2} \sum_{\bk} a_{\alpha}(\bk) \exp[ i \bk \cdot \bm{r}_j]$, 
 Eq.~(\ref{eq:Hij}) is rewritten as 
%------------- (2) ------------------
\begin{eqnarray}
H &=& \sum_{\bk}  \sum_{\alpha, \beta = 1}^4
 t_{\alpha, \beta}(\bk)  a^{\dagger}_{\alpha}(\bk) a_{\beta}(\bk) 
   \nonumber \\
 &=& \sum_{\bk} \sum_{n=1}^4 E_n(\bk) b_n(\bk)^{\dagger} b_{n}((\bk) \; ,
\label{eq:Halphabeta}
\end{eqnarray}
where $\bk = (k_x,k_y)$ and the lattice constant is taken as unity. 
The matrix elements $t_{\alpha,\beta}(\bk)$ 
  $(\alpha, \beta = 1, \cdots 4)$
 are given by 
$t_{12}(\bk) = g_3 + g_2 Y$, 
$t_{13}(\bk) = h_3 + h_2 \bar{X}$, 
$t_{14}(\bk) = h_4 Y + h_1 \bar{X}Y$, 
$t_{23}(\bk) = h_2 + h_3 \bar{X}$, 
$t_{24}(\bk) = h_1 + h_4 \bar{X}$, 
$t_{34}(\bk) = g_1 + g_1 \bar{Y}$, and
$t_{ij}(\bk) = t_{ji}^*(\bk)$, $t_{ii}(\bk) = 0$, 
 where   
$X=\exp[i k_x] = \bar{X}^*$  and  $Y= \exp[i k_y] = \bar{Y}^*$, 
 and the choice of the phase is the same as that of Ref.~\citen{a6}. 
 The eigenvalue, i.e., the energy $E_{n}(\bk)$ [$E_1(\bm{k})> E_2(\bm{k})> E_3(\bm{k}) > E_4(\bm{k})$] 
 is calculated  by diagonalizing the 4 $\times$ 4 matrix Hamiltonian, 
 where $b_n$ denotes an operator corresponding to $E_{n}(\bk)$.

%========  Fig 4 ================
%-----------   fig4 ---------------
Figure \ref{fig4} shows $E_1(\bk)$ and $E_2(\bk)$ 
 on the plane of the 1st Brillouin zone.
 There are two Dirac points with  
$E_1(\bkD) = E_2(\bkD) = \ep_{\rm D}$, which are given by 
$\bkD/\pi -(0,1) = (k_x/\pi, k_y/\pi -1) = \pm (0.61,-0.18)$. 
 Hereafter, we take the chemical potential  (= 0.3995 eV  in the original Hamiltonian) as the origin of the energy, 
  which coincides with $\ep_{\rm D}$ due to the 3/4-filled band. 
A pair of Dirac points are symmetric with respect to 
 the  Y $(= 0, \pi)$ 
  point.
The difference between 
 $E_1(Y) (>0)$ and $E_2(Y) (<0)$ 
     is  smaller than  those of 
    the other TRIMs. In fact, 
 their  energies of  [$E_1(\bm{G}/2)$, $E_2(\bm{G}/2)$]  are obtained as 
 (0.2484,   -0.4037), 
 (0.0916,   -0.2137),   
 (0.0340,   -0.0566), and 
 (0.0663,   -0.0468) 
for $\Gamma=(0,0)$, X = $(\pi,0)$, Y= $(0, \pi)$, and M = $(\pi,\pi)$, 
 respectively.   
 The zero-gap state is obtained 
 because   $E_1(\bk) \ge \ep_{\rm D} \ge E_2(\bk).$
%The dip in $E_2(\bk)$ corresponds to another  Dirac point, 
% which  forms a pair of Dirac cone with that of $E_3(\bk)$. 
%--------------------------------------------

%========  Fig 5 ================
%-------------  Fig 5 ------------------
Figure \ref{fig5}(a)  shows  a contour plot of  the energy difference 
 $ \Delta = E_1(\bk)-E_2(\bk)$,  
 where the Dirac point manifests itself  
    at the center of the orange (bright) region
     ($0 < \Delta < 0.03$).
Two Dirac points $\bkD$  shown by  (I) and (II) are symmetric  
with respect to the Y point. 
The orange region is elongated  toward the Y-point.
 The contour for the fixed (small) $\Delta$ shows an  ellipse, 
 where the ratio of major and minor  axes is about 2.4.

Figure \ref{fig5}(b)  shows the band energy with  Dirac cones 
 around the Dirac point (I), where the linear dispersion is verified.
 In terms of the velocities  $v_x$ and $v_y$ ($v_x > v_y$) corresponding to 
 the minor and major axes of the ellipse, respectively,  
 the band energy $E_j$ ($j$ = 1,2) for the conduction and valence bands 
 may be written as 
%----------- (3)------------------
\begin{eqnarray}
E_j(\bq) = v_t q_t - (-1)^j \sqrt{(v_xq_x)^2 + (v_yq_y)^2} \; , 
\label{eq:eq3}
\end{eqnarray}
 where $\bm{q} = \bk -\bkD$. 
The quantity 
 $q_x (q_y)$ denotes the momentum for the minor (major) axis of 
 Fig.~\ref{fig5}(a), where $q_y$ is parallel to 
the elongated one. The quantities $v_t$ and $q_t$ denote the velocity 
and momentum 
 for the tilt of the cone, rerspectively, where the tilt occurs almost along the $q_x$ direction and  $v_t/v_x \sim 0.2$ is
  much smaller than that of \ET.~\cite{Kajita_JPSJ2014} 
From Fig.~\ref{fig5}(a) and the contour of $E_j(\bk)$, ($j$ = 1, 2),  it is found that 
  the Dirac cone with a linear dispersion may be valid 
 for $E_1(\bk) - E_2(\bk) < 0.05$.

%-----------  parity -----------
 Now,  we examine the presence of Dirac points  by calculating the parity 
 of the eigenfunction 
    at the TRIMs
%, which are given by} $\bm{k}= \bm{G}/2$ 
%      with $\bm{G}$ being the reciprocal lattice vector, 
 as shown in 
 Fig.~\ref{fig5}(a). 
The parity is calculated using the inversion  matrix,~\cite{Piechon2013} 
which is a $\pi$-rotation around  the inversion center
 of the midpoint of  A and A' [closed circle in Fig.~\ref{fig2}(b)]. 
The  4 $\times$ 4 inversion matrix 
 $\hat{P}_i$  has only the diagonal elements given by 
   $1, -1, e^{- i k_x}$ and  $e^{- i k_x - i k_y}$. 
Using $[\hat{H}(\bm{G}/2), \hat{P}_i(\bm{G}/2)]=0$ 
 and  the transformation   of the base~\cite{Piechon2013b}  
 [$u_1 = 2^{-1/2}(u_A+u_{A'})$ and 
 $u_2= -i2^{-1/2} (u_A-u_{A'})$],  
we obtain
 $ \hat{P}_i(\bm{G}/2) \Psi_j(\bm{G}/2)$
  = $E_P(j,\bm{G}/2) \Psi_j(\bm{G}/2)$, where  
 $E_P(j,\bm{G}/2) = \pm 1$ denotes the parity eigenvalue, and 
 $\Psi_j(\bm{G}/2)$ is the eigenfunction for $E_j(\bm{G}/2)$.
This parity for \stfb is summarized in Table \ref{table_1}, 
which satisfies the condition 
for the existence of the  Dirac points,~\cite{Fu2007_PRB76,Piechon2013} 
%-----------   (4) ---------------------
\begin{eqnarray}
 P  = E_P(1,\Gamma) E_P(1,X) E_P(1,Y) E_P(1,M) = -1 \; . 
\label{def:Dirac}
\end{eqnarray}
Note that Eq.~(\ref{def:Dirac}) becomes  +1 
 when the insulating state is expected, i.e., the absence of Dirac points 
 after  merging  by the level crossing between 
  $E_1$ and  $E_2$.  

%------------- Table  1------------------
\begin{table}
\caption{ 
Parity eigenvalue $E_{P}(j,\bm{G}/2) (= \pm 1)$ for  eigenfunction $E_j(\bm{G}/2)$ }
\begin{center}
\begin{tabular} {ccccc}
\hline\noalign{\smallskip}
 & $\Gamma$ & X & Y & M \\
\noalign{\smallskip}\hline\noalign{\smallskip}
%----------------------------------------------
$E_1$ & 1  & -1  & -1 & -1  \\
$E_2$ & 1  & 1 & 1 & 1 \\
$E_3$ & -1  & -1 & -1 & -1 \\
$E_4$ & 1 & -1 & 1 & 1 \\
%------------------------------------------
\noalign{\smallskip}\hline
\end{tabular}
\end{center}
\label{table_1}
\end{table}

%-
%----------   relevance to the experiment ---------------------- 
We also discuss the relevance of the calculated band structure
 to the experiment
on the conductivity of \stf, 
 which  suggests the zero-gap  behavior under ambient pressure. 
% or  under much lower pressure  than \ET.
The experimental finding    
 of  the  maximum  conductivity in addition to 
   the anisotropy of the cone being larger than that of \ET~\cite{Tajima_private}
 is of interest to justify the present Dirac cone.  
Our  calculation shows  that 
 $\sigma_x/\sigma_y$ $\simeq (v_x/v_y)^2$ $\simeq  5$~\cite{Suzumura2014_JPSJ},  whereas the direction of  the  maximum conductivity 
  is sensitive to the choice of transfer energies, 
 and the comparison with the experiment  remains 
 as  a future problem.  

In addition to Eq.~(\ref{eq:Hij}), we briefly mention the effect of 
  the local  potentials $V_B$ and $V_C$  on   
 the B and C sites, which has been examined  for \ET \;\; in Ref.~\citen{Kondo2009_JPSJ}.
 The present calculation for $|V_B| <0.2$, $|V_C| < 0$
 shows that 
 the zero-gap state  exists in region (I) $V_B \leq 0$ and $V_C \leq 0$
 and  that the boundary followed by  merging  
  between the zero-gap state and the insulating state   
    is located slightly outside of  region (I), whereas  
  the metallic region as found in Ref.~\citen{Kondo2009_JPSJ} 
 is absent owing to  the lack of  overtilted Dirac cones.

In summary, we have reported on the following two subjects. (1) We have evaluated the transfer integrals of \stfb for the tight-binding model
 based on our proposal of approximation treating the disorder at the STF sites.
 (2) Analyzing the model in terms of parity at the TRIMs, 
 we found the zero-gap state with Dirac cones. 
% where the Dirac points are close to  merging at the Y point. 
%The finding (1) has the experimental ground regarding the X-ray structural analysis and the conductivity data (Fig.~\ref{fig1}). 
 Such Dirac cones  obtained from (1) and (2) 
   could be justified by  the measurement of  anisotropic conductivity, 
% which is the most fundamental experiment. 
which belongs to the most fundamental and practical experiments.

%========================================

%=========================================================================
\acknowledgements
%=========================================================================
The authors acknowledge helpful discussion with Naoya Tajima at Toho University.

%========================================

%========================================
\end{document}